\documentclass[aps,prb,twocolumn,showpacs]{revtex4}
\usepackage{graphicx}% Include figure files
\usepackage{dcolumn}% Align table columns on decimal point
\usepackage{bm}% bold math
\usepackage[colorlinks,plainpages=false,linkcolor=black,urlcolor=blue,citecolor=blue,pdfpagemode=UseNone,pdfstartview=FitH]{hyperref}
\usepackage{hyperref}
\usepackage{amsmath,bm}
\usepackage{amssymb}

\begin{document}

%%%%%%%%%%%%%%%%%%%%%%%%%%%%%%%%%%%%%%%%%%%%%%%%%%%%%%%%%%%%%%%%%%%%%%%%%%%%%%%%%%%%%%%%%%%%%%%%%%%%%%%%%%%%%%%%%%%%%%%%%%%%%%%%%%%
\title{Evidence for conventional superconductivity in SrPd$_{2}$Ge$_{2}$}
%from combined ARPES, STS and LDA studies}
%
\author{T.~K.~Kim$^{1,2}$
\footnote[1]{Electronic address: \href{mailto:timur.kim@diamond.ac.uk}{timur.kim@diamond.ac.uk}},
A.~N.~Yaresko$^{3}$,
V.~B.~Zabolotnyy$^{2}$,
A.~A.~Kordyuk$^{2,4}$,
D.~V.~Evtushinsky$^{2}$,
N.~H.~Sung$^{5}$,
B.~K.~Cho$^{5,6}$,
T.~Samuely$^{7}$,
P.~Szab\'{o}$^{8}$,
J.~G.~Rodrigo$^{9}$,
J.~T.~Park$^{3}$,
D.~S. Inosov$^{3}$,
P.~Samuely$^{8}$,
B.~B\"{u}chner$^{2}$, and
S.~V.~Borisenko$^{2}$}
\affiliation{
\mbox{$^{1}$ Diamond Light Source Ltd., Didcot, Oxfordshire, OX11 0DE, United Kingdom}\\
\mbox{$^{2}$ Leibniz-Institute for Solid State and Materials Research Dresden, P.O.Box 270116, D-01171 Dresden, Germany}\\
\mbox{$^{3}$ Max-Planck-Institute for Solid State Research, Heisenbergstra$\beta$e 1, D-70569 Stuttgart, Germany}\\
\mbox{$^{4}$ Institute of Metal Physics of National Academy of Sciences of Ukraine, 03142 Kyiv, Ukraine}\\
\mbox{$^{5}$ School of Materials Science and Engineering, Gwangju Institute of Science and Technology, Gwangju 500-712, Korea}\\
\mbox{$^{6}$ Department of Nanobio Materials and Electronics, Gwangju Institute of Science and Technology, Gwangju 500-712, Korea}\\
\mbox{$^{7}$ Institute of Physic at the Faculty of Science, P. J. \v{S}af\'{a}rik University, Park Angelinum 9, 04001 Ko\v{s}ice, Slovakia}\\
\mbox{$^{8}$ Centre of Low Temperature Physics at the Institute of Experimental Physics,}
\mbox{Slovak Academy of Sciences, Watsonova 47, 04001 Ko\v{s}ice, Slovakia}\\
\mbox{$^{9}$ Laboratorio de Bajas Temperaturas, Dept. de F\'{i}sica de la Materia Condensada,} \mbox{Universidad Aut\'{o}noma de Madrid, 28049 Madrid, Spain}\\
}

%%%%%%%%%%%%%%%%%%%%%%%%%%%%%%%%%%%%%%%%%%%%%%%%%%%%%%%%%%%%%%%%%%%%%%%%%%%%%%%%%%%%%%%%%%%%%%%%%%%%%%%%%%%%%%%%%%%%%%%%%%%%%%%%%%%
\date{\today}

\begin{abstract}
Electronic structure of SrPd$_2$Ge$_2$ single crystals is studied by angle-resolved photoemission spectroscopy (ARPES), scanning tunneling spectroscopy (STS) and band-structure calculations within the local-density approximation (LDA).
The STS measurements show single $s$-wave superconducting energy gap $\Delta(0)$ = 0.5\,meV.
Photon-energy dependence of the observed Fermi surface reveals a strongly three-dimensional character of the corresponding electronic bands.
By comparing the experimentally measured and calculated Fermi velocities a renormalization factor of 0.95 is obtained, which is much smaller than typical values reported in Fe-based superconductors.
We ascribe such an unusually low band renormalization to the different orbital character of the conduction electrons and using ARPES and STS data argue that SrPd$_2$Ge$_2$ is likely to be a conventional superconductor, which makes it clearly distinct from isostructural iron pnictide superconductors of the ``122'' family.
\end{abstract}
\pacs{
71.18.+y 	%Fermi surface: calculations and measurements; effective mass, g factor
71.20.-b 	%Electron density of states and band structure of crystalline solids
74.25.Jb 	%Electronic structure (photoemission, etc.)
74.70.Xa 	%Pnictides and chalcogenides
}
\maketitle

\section{Introduction}
Since the discovery of superconductivity in iron pnictides,~\cite{Kamihara_JACS_2008} several families of these novel superconductors have been studied.
Among them, a broad family of the so-called ``122'' superconductors based on $A$Fe$_2$As$_2$ systems ($A$ = Ca, Sr or Ba) with transition temperatures up to $T_{\rm c}\approx38$\,K was prepared by a charge carrier doping, i.e. by partial substitution of alkaline metals for alkaline earth metals, or by partial replacement of Fe (in [Fe$_2$As$_2$] layers) with other 3$d$ transition metals, such as Co or Ni,~\cite{Ishida_JPSJ_2009, Paglione_NPhys_2010_Rev} or by partial substitution of As with P.~\cite{ZhiRen_PRL09_EuFeAsP, Jiang_JPCM09_BFAP}
%% cuprates
Similar to the superconducting cuprates,~\cite{Orenstein_Science_cuprates} all these compounds have quasi-two-dimensional crystal structures formed by iron-pnictide layers separated by different buffer layers.
The partially occupied bands from these iron-pnictide layers determine the electronic structure of the materials in the near Fermi level (FL) region, which in its turn determines the superconducting properties.

One of the puzzles of iron based superconductors is the role of magnetism and the effects of chemical and structural tuning on superconducting properties. Thus, the recent discovery~\cite{Fujii_SPG_0} of a new low-temperature ($T_{\rm c}\approx2.7$\,K) stoichiometric superconductor SrPd$_2$Ge$_2$ isostructural with the group of ``122'' iron pnictides appears intriguing not only because this compound is pnictogen- and chalcogen-free, but also because it has the magnetic metal (Fe) completely replaced by the non-magnetic metal (Pd).
It is, therefore, interesting whether SrPd$_2$Ge$_2$ starts a new family of exotic superconductors similar to pnictides.
In this paper we show that SrPd$_2$Ge$_2$ is, in fact, very different from ``122'' pnictides:
its electronic structure is strongly three-dimensional (3D) and is well described by LDA, it has single isotropic superconducting gap with ${2\Delta}/{kT_{\rm c}}$ not largely exceeding Bardeen-Cooper-Schrieffer (BCS) theory universal value, and its Ginzburg-Landau parameter $\kappa$ calculated from experimental data points to a type-I superconductor, leaving no space for exotic electronic states.

\section{Methods}
Single crystals of SrPd$_2$Ge$_2$ were grown by high temperature flux method using PdGe self flux as described in Ref.~\onlinecite{Sung_SPG_growth}.

\begin{figure}[t!]
\includegraphics[width=\columnwidth]{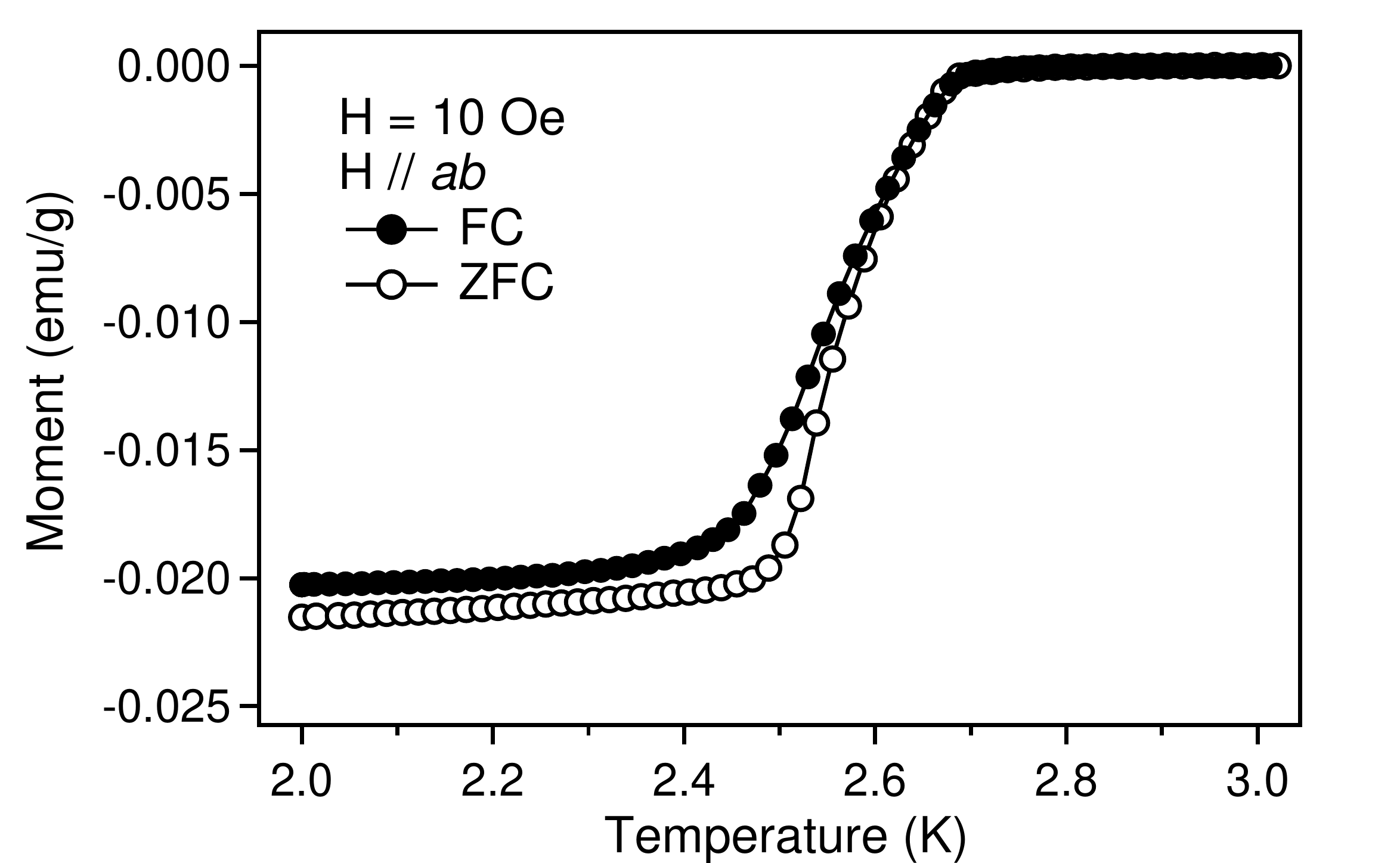}
\caption{\label{Fig_SPG_Magnetization}
Temperature dependence of magnetization of SrPd$_2$Ge$_2$ single crystal with a field of 10\,Oe, perpendicular to the $c$-axis, in both field cooling (FC) and zero-field cooling (ZFC) modes.}
\end{figure}

Temperature- and field-dependent magnetizations of single crystals were measured by a Quantum Design SQUID magnetometer.
Zero-field cooling (ZFC) magnetization was measured with increasing temperature in a field $H$ = 10\,Oe along the $ab$ plane of the sample after cooling down to 2\,K in zero field, and the field cooling (FC) magnetization was measured with increasing temperature in the same field. For the single crystal with a mass of 4 mg the superconducting transition temperature was found to be $T_{\rm c}\approx2.7$\,K (Fig.~\ref{Fig_SPG_Magnetization}).

%STS
Scanning tunneling spectroscopy measurements were done using a homemade low-temperature STM head developed in Ko\v{s}ice in collaboration with UAM Madrid~\cite{Rodrigo_STS_system} and installed in a commercial Janis SSV cryomagnetic system with $^{3}$He-refrigerator and controlled by Nanotec's  Dulcinea SPM electronics.
The atomic-size sharp superconducting tip made of pure lead was scanned over the SrPd$_2$Ge$_2$ sample with bias voltage applied to the tip, while the sample was grounded.

Photoemission experiments were performed at the $1^{3}$-ARPES setup at BESSY based on the R4000 Scienta electron-energy analyzer.~\cite{Borisenko_1cubed}
The geometry of the experiments included fixed analyzer and a sample mounted on a $^{3}$He cryomanipulator that enables rotation about the vertical axis.
The entrance slit of the analyzer was vertically aligned, the angle between optical axis of the analyzer lenses and the incident synchrotron beam was $\sim45^{\circ}$.
All spectra have been measured with linear horizontal polarization.
Single-crystalline samples were cleaved $\emph{in situ}$ in ultra high vacuum at 35\,K.
The measurements performed at temperatures around 1\,K, the overall energy and angular resolutions were set to 10\,meV and 0.2$^\circ$, respectively.

Electronic band structure calculations were performed for the experimental crystal structure of SrPd$_2$Ge$_2$ from Ref.~\onlinecite{Fujii_SPG_0} within the local density approximation (LDA) using the linear muffin-tin orbital (LMTO) method.~\cite{And75, PYLMTO}

\section{Results and Discussion}
The crystal structure of SrPd$_2$Ge$_2$ is the same as in BaFe$_2$As$_2$, but its electronic structure is expected to exhibit a much stronger 3D character.~\cite{Shein_FS_calculation}

\begin{figure}[tb!]
\includegraphics[width=\columnwidth]{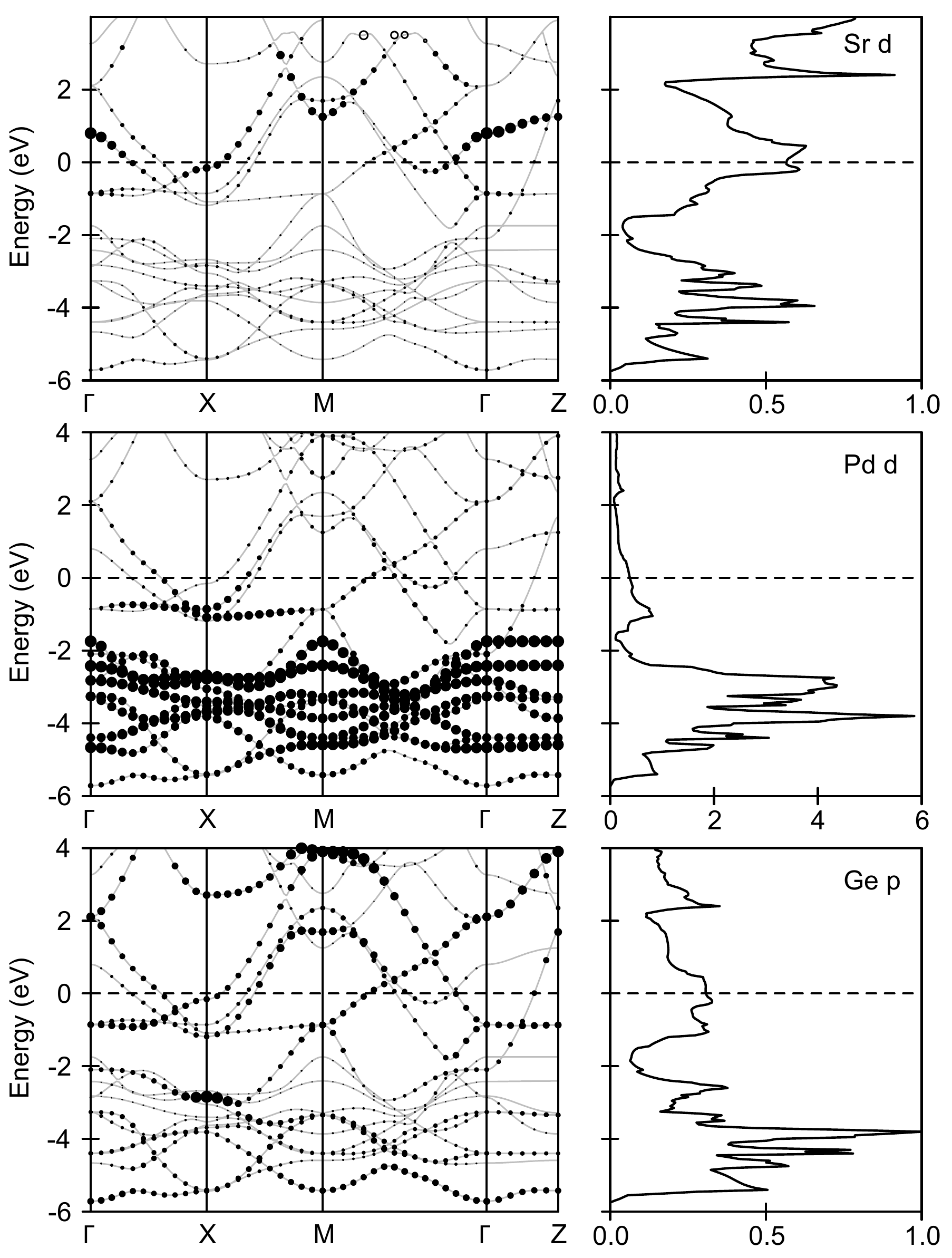}
\caption{\label{Fig_Fat_bands_DOS}
LDA band structure and density of states for SrPd$_2$Ge$_2$. Bands in different panels (from top to bottom) are decorated with circles, whose radii are proportional to the weight of Sr $d$, Pd $d$, and Ge $p$ states in the corresponding Bloch wave functions.}
\end{figure}

\begin{figure}[tb!]
\includegraphics[width=\columnwidth]{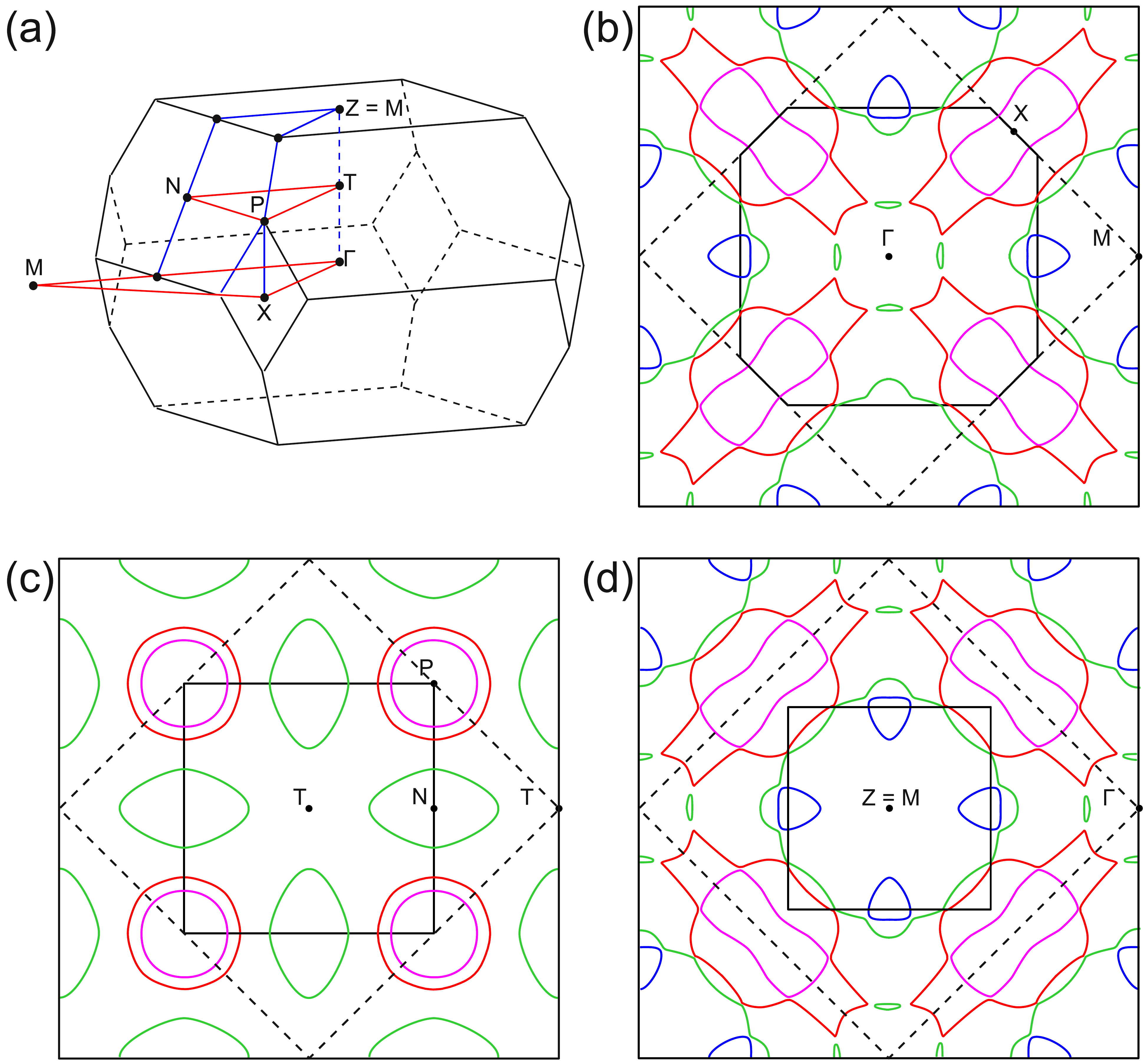}
\caption{\label{Fig_Theory_FS}(Color online)
Brillouin zone of SrPd$_2$Ge$_2$~(a), cuts of the Fermi surface from calculated electronic band structure for different $k_z$ values (b,~c,~d).}
\end{figure}

Our LDA calculations show, that in contrast to the isostructural iron pnictides, in which Fe $d$ states responsible for very peculiar nesting of electron and hole-like sheets of the Fermi surface are partially occupied, the Pd $d$ states in SrPd$_2$Ge$_2$ are completely filled, and
bands crossing the Fermi level are formed by delocalized Ge $p$ and Sr $d$ states with only minor
admixture of the Pd $d$ states (see Fig.~\ref{Fig_Fat_bands_DOS}).
As a result, the calculated Fermi surface (FS) shown in Fig.~\ref{Fig_Theory_FS} reveals a 3D character of the SrPd$_2$Ge$_2$ electronic structure with very strong $k_z$ dependence of the conduction bands.
The calculated two-dimensional Fermi surfaces corresponding to the cuts of the Brillouin zone (BZ) with different $k_z$ values are shown in Fig.~\ref{Fig_Theory_FS} b, c, d.

\begin{figure}[htb]
\includegraphics[width=\columnwidth]{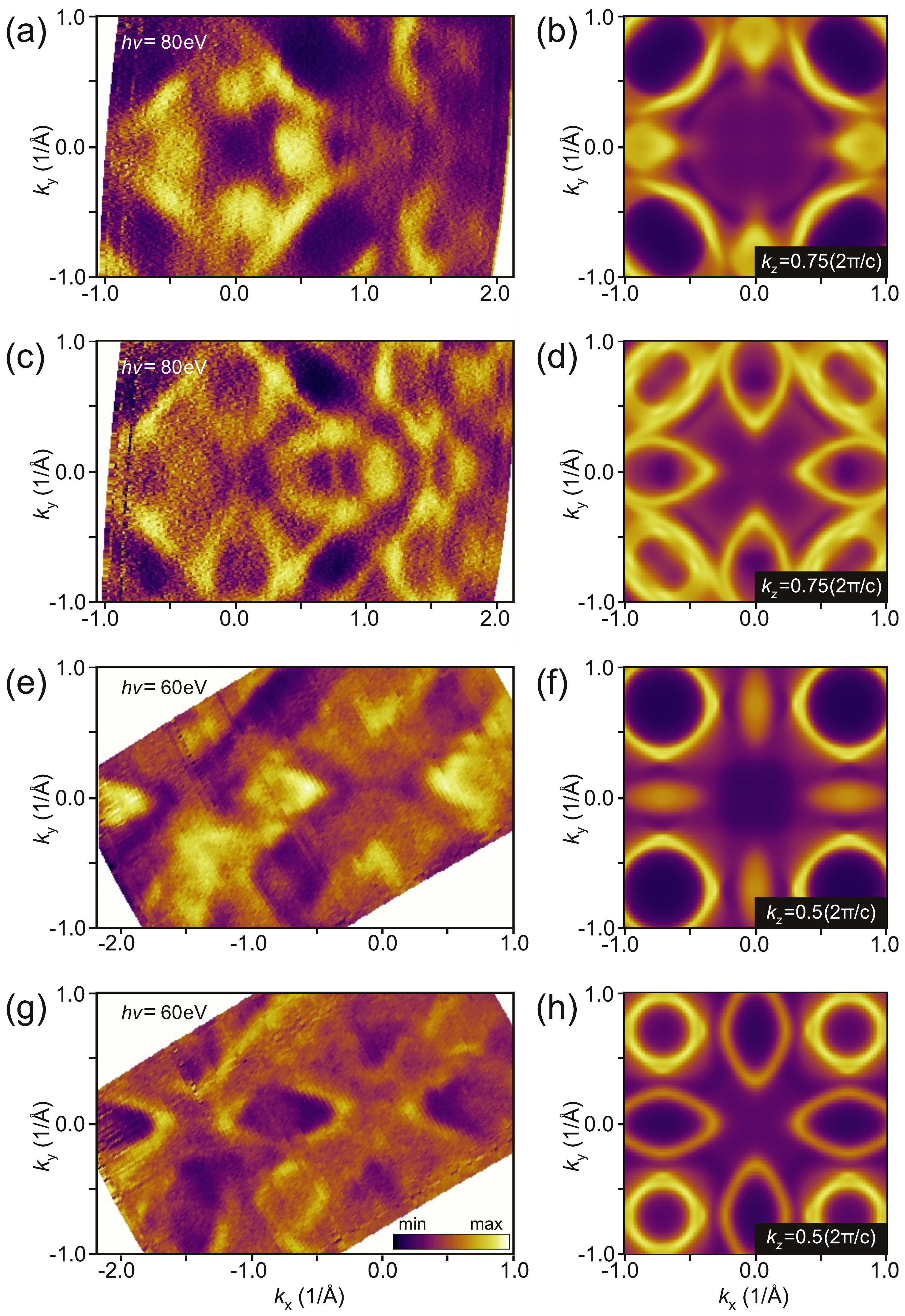}
\caption{\label{Fig_MDMs}(Color online)
Comparison of the experimental and simulated MDMs: Photoemission intensity measured with $h\nu$=80\,eV at the FL (a) and 500\,meV below the FL (c), and corresponding density of states calculated with $k_z$=0.75\,(2$\pi/c$) (b,~d).
Photoemission intensity measured with $h\nu$=60\,eV at the FL (e) and 500\,meV below the FL (g), and corresponding density of states calculated with $k_z$=0.5\,(2$\pi/c$) (f,~h). Sample temperature is 1.3\,K.}
\end{figure}

%FS topology from ARPES
Aiming to examine the topology of the Fermi surface of SrPd$_2$Ge$_2$ experimentally, an angle-resolved photoemission spectroscopy study has been performed over extended area in momentum space.
The momentum distribution maps (MDMs) derived from the ARPES experiment with $h\nu$=80 and 60\,eV at 1.3\,K are shown in Fig.~\ref{Fig_MDMs}.
Indeed, a very strong photon-energy dependence of the ARPES data is observed, indicating a strong $k_z$ dependence of the Fermi surface.

In order to understand the experimentally observed FS topology, the two-dimensional MDMs with different $k_z$ values have been simulated using the calculated electronic structure.
By systematically varying $k_z$ values, we found that the best agreement between experimental and calculated band structures is observed for $h\nu$=80\,eV and $k_z$=0.75\,(2$\pi/c$), both for zero (FL) and 500\,meV binding energies (see Fig.\,\ref{Fig_MDMs} a--d).
Binding energy shift of 460\,meV was applied to LDA band positions to match the $k_{\rm F}$  value of the Fermi level crossing by the electron pocket around the X point.
The same procedure gives the best agreement between the experimental and calculated band structures for $h\nu$=60\,eV  and $k_z$=0.5\,(2$\pi/c$) for both zero (FL) and 500\,meV binding energies (see Fig.\,\ref{Fig_MDMs} e--h).
In both cases, the remarkable agreement between experimentally obtained and calculated band structures is observed.
Therefore by changing the excitation photon energy one could distinguish electronic states with different $k_z$ values with accuracy of at least 0.25 of the vertical BZ size.
This means, assuming that Heisenberg uncertainty principle can be applied to the electronic escape depth and $k_z$ momentum resolution of photoemission experiment, that with conventional ARPES within the ultraviolet photon energy range, the bulk sensitivity of at least 4$\times$$c$~$\approx$~40{\AA} can be achieved.

\begin{figure}[tbh]
\includegraphics[width=\columnwidth]{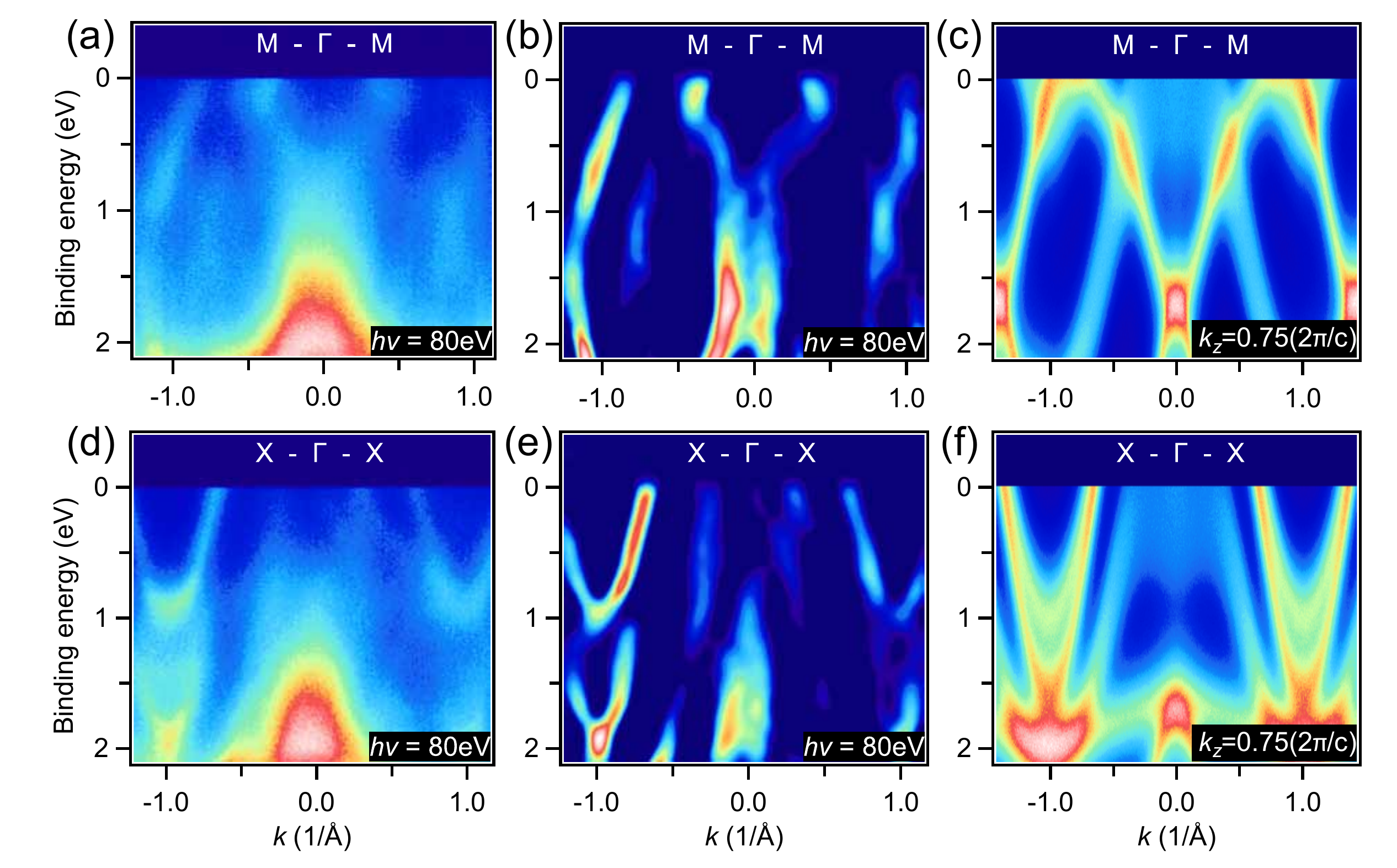}
\caption{\label{Fig_EDMs}(Color online)
Comparison of the experimental and simulated energy distribution maps:
Photoemission intensity measured with  $h\nu$=80\,eV in M-$\Gamma$-M  (a) and X-$\Gamma$-X (d) directions, second derivatives (b) and (e),
and corresponding density of states calculated with $k_z$=0.75\,(2$\pi/c$) (c) and (f). Sample temperature is 1.3\,K.}
\end{figure}

% high symmetry cuts and renormalization determination
Energy distribution maps (EDMs) for BZ cuts in high-symmetry directions for excitation photon energy $h\nu$=80\,eV are shown in Fig.~\ref{Fig_EDMs}. From the comparison of the experimental EDMs with the ones simulated using the calculated electronic band structure, we derive the band renormalization.
%
%band renormalization from band width
For ``122'' iron pnictides LDA calculations usually give overestimated values for the band width, if compared to experimentally derived from ARPES (Table~\ref{table1}).
But in the X-$\Gamma$-X EDM the separation between the bottoms of the two electron pockets around the X point is by $\sim$300\,meV higher as compared to LDA.
Therefore one needs to apply a multiplier of 1.26 to LDA bands to get the best agreement with the experimentally derived band structure. This corresponds to the band renormalization factor of 0.8.

%band renormalization from Fermi velocity
Another approach to obtain the value of renormalization of the band forming electron pockets around the X point is to determine the Fermi velocity of the band at the Fermi level and compare it to the value from the calculated bare band dispersion.
Fitting the positions of momentum distribution curves maxima within first 200\,meV below the Fermi level, we obtain the band dispersion and corresponding value of the Fermi velocity.
The ratio of calculated Fermi velocity $v_{\rm F}^{\rm LDA}$=4.45\,eV{\AA} to experimental Fermi velocity $v_{\rm F}^{\rm ARPES}$=4.7\,eV{\AA} gives the renormalization factor of $\sim$0.95$\pm0.1$.
This value is lower than in iron-based pnictide and chalcogenide superconductors as presented in Table~\ref{table1}.
For example, for the isostructural compound KFe$_2$As$_2$ with a similar $T_{\rm c}$ of 3\,K, the electron band renormalization factor was reported to vary for different bands from 2 to 4.~\cite{Sato_KFe2As2}

% DMFT - correlations in pnictides
This difference in the band renormalization can be explained by the different orbital characters of the Fermi surfaces of SrPd$_2$Ge$_2$ and of the iron pnictides.
In the former, the corresponding bands are dominated by delocalized Ge $p$ and Sr $d$ states for which the effects of electronic correlations are treated well enough already by LDA.
In the latter, on the other hand, the bands crossing the Fermi level are formed by moderately correlated Fe $d$ states.
The importance of the correlations seems to be confirmed by the dynamical mean-field theory calculations which give effective band renormalization values of 2-3 for iron pnictides,~\cite{Skornyakov_DMFT_122, Kutepov_DMFT_122, Haule_DMFT_1111, Laad_DMFT_1111} that are in a good agreement with ARPES data (Table~\ref{table1}).

\begin{table}[bt]
\caption{\label{table1} Bandwidth renormalization factor, $m$$^*$/$m$, and superconducting transition temperature, $T_{\rm c}$, for different iron pnictides.}
\begin{ruledtabular}
\begin{tabular}{llll}
Compound                           & $m$$^*$/$m$ & $T_{\rm c}$, K   & Reference\\
\hline
BaFe$_2$As$_2$                       & 1.5       &  -   & Ref.\onlinecite{Shen_BaFe2As2_ARPES}\\
Ba$_{0.6}$K$_{0.4}$Fe$_2$As$_2$      & 2.7       & 37   & Ref.\onlinecite{Shen_BaFe2As2_ARPES}\\
Ba$_{0.6}$K$_{0.4}$Fe$_2$As$_2$      & 1.3--9    & 37   & Ref.\onlinecite{Ding_BKFA}\\
Ba(Fe$_{0.94}$Co$_{0.06}$)$_2$As$_2$ & 1.7       & 25   & Ref.\onlinecite{Shen_BaFe2As2_ARPES}\\
LiFeAs                               & 3         & 18   & Ref.\onlinecite{Borisenko_LiFeAs}\\
FeTe$_{1-x}$Se$_x$                   & 6--20     & 11.5 & Ref.\onlinecite{Tamai_FeTeSe_ARPES}\\
FeTe$_{1-x}$Se$_x$                   & 3         & 9    & Ref.\onlinecite{Feng_FeTeSe_ARPES}\\
NaFeAs                               & 5.4--6.5  & 8    & Ref.\onlinecite{Feng_NaFeAs_ARPES}\\
LaFePO                               & 2.2       & 5.9  & Ref.\onlinecite{Shen_LaOFeP_Nature}\\
KFe$_2$As$_2$                        & 2--4      & 3    & Ref.\onlinecite{Sato_KFe2As2}\\
SrPd$_2$Ge$_2$                       & 0.8--0.95 & 2.7  & this work\\
\end{tabular}
\end{ruledtabular}
\end{table}

The weaker band mass renormalization, as compared with iron pnictides, together with the strong 3D character of the electronic structure and a non-magnetic ground state suggest that superconductivity in SrPd$_2$Ge$_2$ is conventional and presumably of the electron-phonon nature.
A recent study of the specific heat suggests a strong electron-phonon interaction
in SrPd$_2$Ge$_2$,~\cite{Sung_SPG_growth} however it shows a significant deviation from the
weak-coupling behavior in this material.

%%% ksi lambda
In order to clarify the nature of the superconductivity in SrPd$_2$Ge$_2$, one can use the knowledge of Fermi surface topology, Fermi velocity and energy gap, and estimate values for the coherence length $\xi$ and London penetration depth $\lambda_{\rm L}$.~\cite{Inosov_LiFeAs}
The Ginzburg-Landau parameter $\kappa = \lambda_{\rm L}~/~\xi$, refers to the type of superconductor: type-I superconductors are those with $0 < \kappa < 1/{\sqrt{2}}$, and type-II superconductors are those with $\kappa > 1/{\sqrt{2}}$.

The Cooper pair coherence length $\xi$ and the magnetic field penetration depth $\lambda_{\rm L}$ can be estimated from microscopic parameters of the electronic spectrum as follows:~\cite{Tinkham, Chandrasekhar}
\begin{equation}
\xi = \frac{\hbar v_{\rm F}}{\pi\Delta} \propto \frac{v_{\rm F}}{\Delta},
\label{xi}
\end{equation}

\begin{equation}
\lambda_{\rm L} = \left(\frac{e^2}{4\pi^2\varepsilon_0c^2\hbar L_c} \int v_{\rm F}{\rm d}k\right)^{-\frac{1}{2}} \propto \frac{1}{\sqrt{\langle v_{\rm F} \rangle \cdot \langle l^{\rm 2D}_{\mathbf{k}}\rangle}},
\label{lambda}
\end{equation}
where $\varepsilon_0$, $\hbar$, $e$, $c$ are physical constants, $L_c$ is the $c$-axis lattice parameter, $v_{\rm F}$ is the Fermi velocity, $\langle l^{\rm 2D}_{\mathbf{k}} \rangle$ is the length of the Fermi contours (averaged over different $k_z$ values for the three dimensional case), $\Delta$ is the value of the superconducting gap. %(maximal value is taken in the calculation for certainty).

\begin{figure}[t!]
\includegraphics[width=\columnwidth]{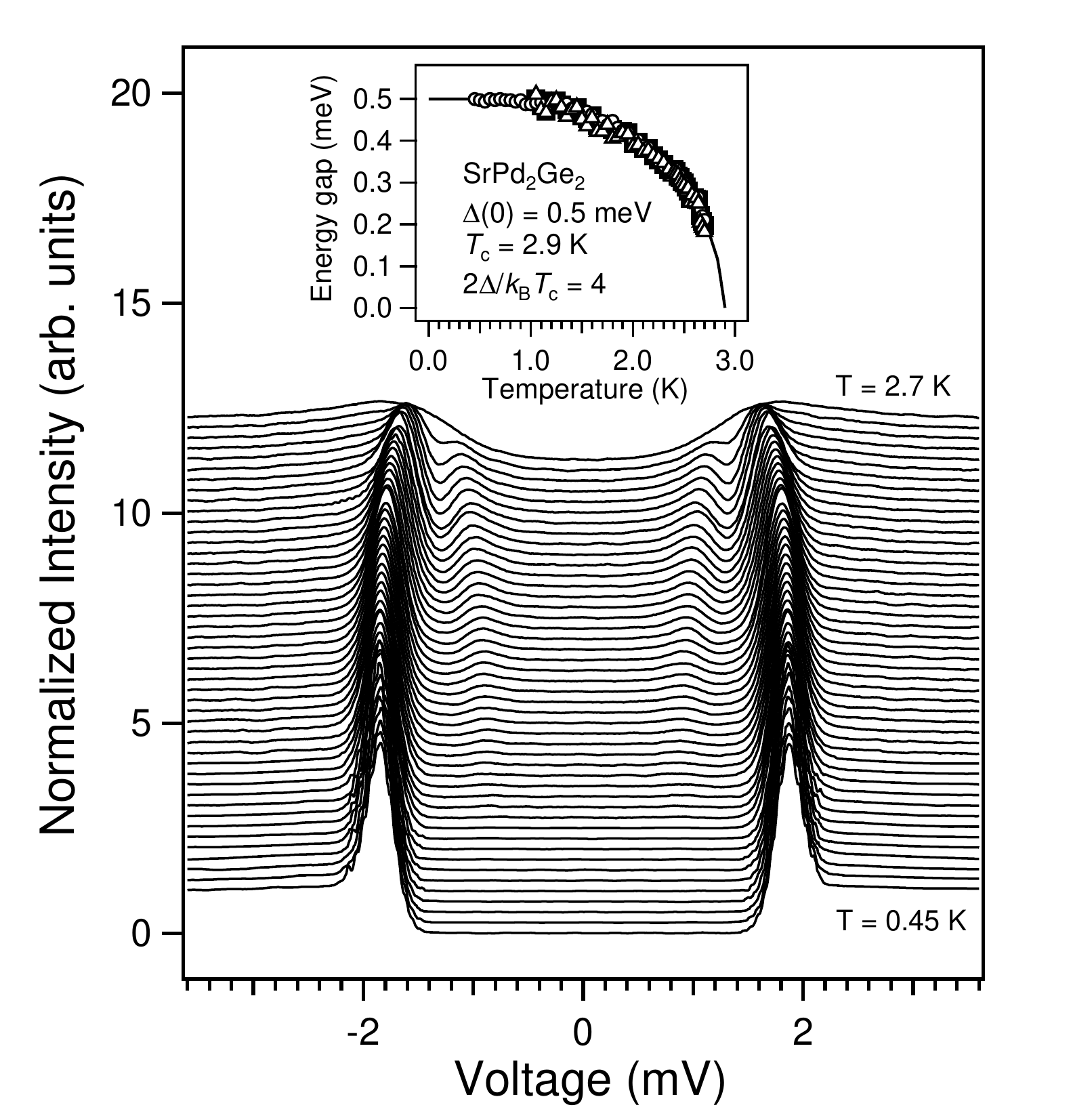}
\caption{\label{Fig_SPG_STS}
STS conductance spectra of the superconductor-superconductor junction, Pb - SrPd$_2$Ge$_2$, measured in zero magnetic field at different temperatures between 0.45\,K (lowest curve) to 2.7\,K, increasing by 0.05\,K (upper curves are shifted for clarity).
Upper inset: Temperature dependence of the superconducting gap of SrPd$_2$Ge$_2$ (circles, squares and triangles) in comparison with the BCS theory (line).}
\end{figure}

%STS data
The superconducting energy gap of SrPd$_2$Ge$_2$ can be directly determined
from low-temperature STS measurements.
Figure~\ref{Fig_SPG_STS} shows the tunneling conductance spectra between the superconducting Pb tip and SrPd$_2$Ge$_2$ sample measured at different temperatures ranging from 0.45\,K to 2.7\,K.
Each of these differential conductance versus voltage spectra is proportional to the convolution of the superconducting density of states of both electrodes forming a junction.
All curves exhibit two large peaks located at ${ca.}$ $\pm$1.86\,mV for the lowest temperatures.
These peaks, corresponding to the sum of the superconducting energy gaps of the tip and the sample, appear at voltages $\pm\vert\Delta_{\rm Pb} + \Delta_{\rm S}\vert/e$, where $\Delta_{\rm Pb}$ and $\Delta_{\rm S}$ are the superconducting energy gaps of the lead tip and the sample, respectively, and $e$ is the electron's charge.
In addition, at temperatures above 1.2~K, two minor peaks at ${ca.}$ $\pm$0.86\,mV come out.
These peaks, corresponding to the difference of the superconducting energy gaps of the tip and the sample, appear at voltages $\pm\vert\Delta_{\rm Pb} - \Delta_{\rm S}\vert/e$, and represent the thermally activated current induced by excited quasiparticles above and corresponding holes below the superconducting energy gap of the sample.
For the curves taken at the lowest temperatures the zero conductance plateau in the center of the respective curve goes up to well above $\Delta_{\rm Pb}$.
As temperature is raised, the dip appearing at ${ca.}$ $\pm$1.3\,mV between the two above
mentioned peaks reaches even negative conductance values.
These two observations indicate that SrPd$_2$Ge$_2$ is indeed an $s$-wave single gap superconductor, and $s^\pm$ pairing proposed for the isostructural ``122'' iron pnictides,~\cite{Teague_PRL11_BFCA_STS_2gaps} which has been associated with unconventional pairing mediated by magnetic fluctuations, is probably absent here.

%STS figure

% SC gap T dep
The two pairs of peaks corresponding to $\vert\Delta_{\rm Pb}+\Delta_{\rm S}\vert$ and $\vert\Delta_{\rm Pb}-\Delta_{\rm S}\vert$ allow a direct determination of $\Delta_{\rm Pb}(T)$ and  $\Delta_{\rm S}(T)$ from the tunneling curves.
The superconducting energy gap value of lead, $\Delta_{\rm Pb}$ = 1.36\,meV, is obtained at the lowest temperature.
It is in perfect agreement with literature.~\cite{Kittel}
The superconducting energy gap value of SrPd$_2$Ge$_2$, $\Delta_{\rm S}(T)$, can then be estimated in three different fashions.
First, by subtracting the values of the two peaks ($\vert\Delta_{\rm Pb}+\Delta_{\rm S}\vert$ - $\vert\Delta_{\rm Pb}-\Delta_{\rm S}\vert$)/2, second, by subtracting $\Delta_{\rm Pb}$ from $\vert\Delta_{\rm Pb}+\Delta_{\rm S}\vert$ and third, by subtracting $\vert\Delta_{\rm Pb}-\Delta_{\rm S}\vert$ from $\Delta_{\rm Pb}$, indicated by black squares, open circles and triangles in the inset of Fig.~\ref{Fig5_SPG_STS}, respectively.
All three estimates of $\Delta_{\rm S}(T)$ coincide accurately with the prediction of the BCS theory.
The resulting superconducting energy gap and critical temperature of SrPd$_2$Ge$_2$, measured by STS, are  $\Delta(0)$ = 0.5\,meV and $T_{\rm c}$ = 2.9\,K, indicating strong coupling superconductivity with a ratio of ${2\Delta}/{kT_{\rm c}}$ = 4.0.

%% GL parameter
Taking the maximum value of the superconducting gap from STS data and details of the Fermi surface topology and Fermi velocity from ARPES data, in Table~\ref{table_xi_lambda}, the formulae~(\ref{xi}, \ref{lambda}) are used to estimate the in-plane Pippard superconducting coherence length and London penetration depth, and consequently to evaluate the Ginzburg-Landau parameter.
%
%% GL table
\begin{table*}[thb!]
\caption{Average Fermi velocity, $v_{\rm F}$, length of the Fermi contours, $\langle l^{\rm 2D}_{\mathbf{k}} \rangle$, and the superconducting gap, $\Delta_{\rm max}$, for (Ba,K)Fe$_2$As$_2$, LiFeAs, Ba(Fe,Co)$_2$As$_2$, SrPd$_2$Ge$_2$.
The London penetration depth, $\lambda_{\rm L}$, and the coherence length, $\xi$, as estimated from aforementioned parameters according to formulae (\ref{xi}, \ref{lambda}). The Ginzburg-Landau parameter, $\kappa$, is shown in the last column.}
\label{table_xi_lambda}
\begin{ruledtabular}
\begin{tabular}{l@{~}l@{~~~}                         l@{~~~}                    l@{~~~}        l@{~~~}   l@{~~~}   l@{~}}%
Compound &$\langle v_{\rm F}\rangle$, eV${\rm \AA}$ & $\langle l^{\rm 2D}_{\mathbf{k}} \rangle$, ${\rm \AA}^{-1}$ & $\Delta_{\rm max}$, meV & $\lambda_{\rm L}$, nm & $\xi$, nm & $\kappa = \lambda_{\rm L}/\xi$\\%
\hline
(Ba,K)Fe$_2$As$_2$    &~~0.41\,\cite{Evtushinsky_NJP}&~~$2\pi\cdot0.74$\,\cite{Evtushinsky_NJP}&~~10\,\cite{Evtushinsky_PRB_2009}&170&1.3&131\\
LiFeAs                &~~0.31\,\cite{Borisenko_LiFeAs,Inosov_LiFeAs}&~~$2\pi\cdot0.96$\,\cite{Borisenko_LiFeAs}&~~3\,\cite{Inosov_LiFeAs}&172&3.2&54\\
Ba(Fe,Co)$_2$As$_2$   &~~0.7\,\cite{Co122}&~~$2\pi\cdot0.61$&~~5&144&4.3&33\\
SrPd$_2$Ge$_2$        &$\sim4.7$&~~$\sim2\pi\cdot1.2$&$0.5$&40&291&0.14\\
\end{tabular}
\end{ruledtabular}
\end{table*}
If for iron-pnictide superconductors in Table~\ref{table_xi_lambda} the obtained Ginzburg-Landau parameter $\kappa \gg 1/{\sqrt{2}}$ indicates that these materials are type-II superconductors, then for SrPd$_2$Ge$_2$ the obtained $\kappa < 1/{\sqrt{2}}$ indeed points to a type-I superconductor.
% weak-strong coupling
Therefore, the fact that SrPd$_2$Ge$_2$ is isostructural to ``122'' family of iron pnictides does not necessarily lead to the same origin of the superconductivity.
As has been recently suggested,~\cite{Inosov_coupling} even iron pnictides within a single family may not necessarily share the same superconducting pairing mechanism.
This is best demonstrated, for example, by the presence of unconventional superconductivity in Ba(Fe$_{1-x}$Ni$_x$)$_2$As$_2$ close to optimal doping ($x \approx 0.05$)~\cite{Li_Ni_2009, Chi_Ni_2009} and the conventional phonon-mediated pairing in BaNi$_2$As$_2$ ($x=1$).~\cite{Kurita_Ni_2011}

\section{Summary}
In conclusion, the occupied electronic structure of the pnictogen-free SrPd$_2$Ge$_2$ has been studied for the first time by means of ARPES and compared with first-principles calculations.
At variance with isostructural iron-based superconductors, its electronic structure reveals a much more pronounced three-dimensional character.
3D structure of SrPd$_2$Ge$_2$ Fermi surface is confirmed by the remarkable agreement of LDA calculations with experimentally measured momentum distribution maps.
In contrast to iron-based superconductors, the orbital composition of the conductance band is not dominated by the transition-metal $d$ states, which are localized much deeper below the Fermi level, but represents a mixture of Sr $d$, Pd $d$, and Ge $p$ states.

By comparing the experimental and calculated band structures, the values of the out-of-plane component of the electron momentum corresponding to the photoemission spectra obtained with different excitation photon energies has been determined, indicating bulk sensitivity of the ARPES method of the order of 4 unit-cell layers along the $c$ direction.

Using the ratio of the calculated bare Fermi velocity to the experimental one, the band renormalization factor of $\sim$0.95 has been obtained.
This relatively small value of electron band renormalization together with a relatively low $T_c$ as compared to iron pnictides and chalcogenides support the conventional, phonon-mediated mechanism of superconductivity in this pnictogen-free compound.

The STS measurements show that SrPd$_2$Ge$_2$ is a strong coupling single $s$-wave gap superconductor, with superconducting energy gap $\Delta(0)$ = 0.5\,meV and the BCS-like temperature dependence of the gap.
Moreover, the estimation for the Ginzburg-Landau parameter $\kappa =0.14$ obtained from ARPES and STS data indicates that SrPd$_2$Ge$_2$ is likely to be a type-I superconductor, contrary to the conclusions from magnetization studies.~\cite{Sung_SPG_growth}
Further investigations by means of STM are in progress and will be published elsewhere.~\cite{Samuley_LT-STM}
%%%  END

\begin{acknowledgments}
This work was supported by the DFG priority program SPP1458, Grants No. KN393/4, BO1912/2-1, BO3537/1-1 (DSI, JTP);
Slovak  Research and Development  Agency under the contract No. VVCE-0058-07, the Slovak grant VEGA no. 0148/10 and 1/0138/10, 7th FP MNT - ERA.Net II. ESO (TS, PS, JGR, PS);
and by the Korea government (MEST) Grant No. R15-2008-006-01002-0 (NHS, BKC).

\end{acknowledgments}


\begin{thebibliography}{99}
\bibitem{Co122}
The extraction of the electronic band parameters for Co-doped 122 iron arsenide Ba(Fe,Co)$_2$As$_2$ from ARPES data will presented elsewhere.
\end{thebibliography}
\end{document}